%% file: main.tex
\documentclass[runningheads]{llncs}

\usepackage[utf8]{inputenc}
\usepackage[T1]{fontenc}

\usepackage{amsmath}
\usepackage{amssymb}
\usepackage{bm}

\usepackage{graphicx}
\usepackage{wrapfig}
\usepackage{float}
\usepackage[skip=3pt]{caption}

\usepackage{array}
\usepackage{tabularx}
\newcolumntype{C}{>{\centering\arraybackslash}X}
\usepackage{booktabs}
\usepackage{multirow}
\usepackage{threeparttable}
\usepackage{makecell}
\usepackage{siunitx}
\usepackage[table]{xcolor}
\definecolor{crimson}{RGB}{220,20,60}
\definecolor{forestgreen}{rgb}{0.13,0.55,0.13}
\definecolor{headerblue}{HTML}{D9EEF7}
\definecolor{catchup}{HTML}{C6E6A9}
\definecolor{catchupEM}{HTML}{FDE2A7}
\definecolor{catchupBLEU}{HTML}{D7C7FF}
\usepackage{soul}

\usepackage{tikz}
\usetikzlibrary{plotmarks}
\usepackage{pgfplots}
\pgfplotsset{compat=1.18}

\usepackage{fvextra}
\usepackage{tcolorbox}
\tcbuselibrary{breakable,skins,listings}
\newtcblisting{promptbox}{
  enhanced,
  breakable,
  colback=black!5!white,
  colframe=black!80!white,
  listing engine=listings,
  listing only,
  fontupper=\ttfamily\small,
  listing options={
    breaklines=True,
    tabsize=2,
    showspaces=false, showtabs=false
  }
}

\usepackage{hyperref}

\begin{document}
\title{Less LLM, More Documents:\\
Searching for Improved RAG}

\author{%
Jingjie Ning$^{*}$ \and
Yibo Kong$^{*}$ \and
Yunfan Long$^{*}$ \and
Jamie Callan
}
\authorrunning{J. Ning et al.}
\institute{School of Computer Science, Carnegie Mellon University, Pittsburgh, PA, USA \email{\{jening,yibok,justinlo\}@andrew.cmu.edu}, \email{callan@cs.cmu.edu}}
\renewcommand{\thefootnote}{\fnsymbol{footnote}}
\setlength{\skip\footins}{20pt}
\footnotetext[1]{Equal Contributions.}


%
\maketitle              
\input{0-abstract}

\input{1-introduction}

\input{2-related-work}

\input{3-methodology}

\input{4-Experiment}

\section{Results \& Analysis}
\input{5.1-corpus-compensation}

\input{5.2-llm-sizes}

\input{6-conclusion}

\newpage
\appendix
\input{appendix}

\newpage
\bibliographystyle{splncs04}
\bibliography{refs}

\end{document}

%% file: 0-abstract.tex
\begin{abstract}
Retrieval-Augmented Generation (RAG) couples document retrieval with large language models (LLMs). While scaling generators often improves accuracy, it also increases inference and deployment overhead. We study an orthogonal axis: enlarging the retriever’s corpus, and how it trades off with generator scale. Across multiple open-domain QA benchmarks, corpus scaling consistently strengthens RAG and can in many cases match the gains of moving to a larger model tier, though with diminishing returns at larger scales. Small- and mid-sized generators paired with larger corpora often rival much larger models with smaller corpora; mid-sized models tend to gain the most, while tiny and very large models benefit less. Our analysis suggests that these improvements arise primarily from increased coverage of answer-bearing passages, while utilization efficiency remains largely unchanged. Overall, our results characterize a corpus--generator trade-off in RAG and provide empirical guidance on how corpus scale and model capacity interact in this setting.

\keywords{Retrieval-Augmented Generation\and Passage Retrieval \and Large Language Models \and Corpus Scaling \and Resource-Constrained Inference}
\end{abstract}

%% file: 1-introduction.tex
\section{Introduction}
Retrieval-Augmented Generation (RAG) \cite{gao2024RAG,rag} augments large language models (LLMs) with retrieved evidence and is widely used for open-domain question answering (QA) \cite{gupta2024comprehensivesurveyretrievalaugmentedgeneration,upadhyay2024QA,voorhees2000trec}. While much prior work improves RAG by scaling the generator, larger models typically incur higher inference and deployment overhead \cite{ghorbani2022scaling,narayanan2021GPU,patterson2021carbonemissionslargeneural}. In parallel, the retriever determines what evidence is available to the generator and can help mitigate hallucinations \cite{rag,shi-etal-2024-replug}, yet how retriever-side scaling interacts with generator scale is not well understood.

We study this interaction by varying the retrieval corpus scale while holding the retrieval pipeline and evidence budget fixed. Concretely, we combine randomly-shardable retrieval over ClueWeb22 \cite{clueweb22} with open-source Qwen3 models \cite{qwen3} spanning 0.6B--14B parameters, and evaluate on NQ, TriviaQA, and WebQ \cite{berant-etal-2013-semantic,joshi2017triviaqa,kwiatkowski2019natural}. Our results show that enlarging the corpus consistently improves RAG and can, in this setting, partially offset smaller generators: for example, a 1.7B model with a $4\times$ larger corpus outperforms a 4B model, and a 4B model with a $2\times$ larger corpus consistently outperforms an 8B model.

Overall, our work characterizes a corpus--generator trade-off in RAG and provides empirical guidance on (i) when corpus scaling can match gains from scaling the generator, (ii) which model-size regimes benefit most from additional corpus, and (iii) why these gains arise, via analyses of answer-bearing evidence coverage and context utilization.

%% file: 2-related-work.tex
\section{Related Work}

A substantial body of research has focused on enhancing the intrinsic capabilities of LLMs. Instruction tuning \cite{wang2022selfinstruct,zhang2023instruction} and prompt engineering \cite{he-etal-2024-position,reynolds2021prompt} improve alignment with user queries, while scaling model size generally yields higher accuracy across diverse tasks \cite{brown2020languagemodelsfewshotlearners,kaplan2020scalinglawsneurallanguage}. However, ever-larger LLMs (e.g., PaLM with 540B parameters \cite{chowdhery2022palm}) incur prohibitive computational costs, which limits their practicality in many settings.

In parallel, retrieval-augmented models have emerged as an alternative path to scaling. Retrieval-augmented language models (RALMs) such as RETRO \cite{pmlr-v162-borgeaud22a} and Atlas \cite{atlas2023} demonstrate that enlarging inference-time datastores consistently improves performance: relatively small generators paired with massive retrieval memory can outperform much larger LM-only baselines. Shao et al. \cite{shao2024scaling} further confirmed this monotonic trend. Unlike modular RAG, which decouples retriever and generator, RALMs integrate retrieval through pretraining-time vector memories, requiring retriever-generator co-training. In contrast, we focus on modular RAG with a fixed evidence budget and study corpus–generator trade-offs across model-size regimes, rather than monotonic retrieval scaling in integrated RALMs.

Modular RAG instead relies on external corpora that can be scaled independently of the generator. This line of work has progressed from Dense Passage Retrieval (DPR) \cite{karpukhin-etal-2020-dense} to efficiency-oriented methods such as ANCE \cite{xiong2021approximate} and Contriever \cite{izacard2021contriever}, which make retrieval over very large corpora feasible. These advances enabled a shift from early Wikipedia-only setups to broader and more diverse corpora such as LoTTE \cite{santhanam-etal-2022-colbertv2} and BEIR \cite{thakur2021beir}. However, while the community has implicitly moved toward increasingly larger corpora, the direct impact of corpus growth itself has not yet been systematically and comprehensively examined.

More recently, studies have begun to explore broader factors influencing RAG, including model size, corpus scale, and context size \cite{li-etal-2025-enhancing-retrieval,vladika2025size}. However, these analyses remain fragmented and primarily descriptive, typically isolating single variables. Importantly, they do not provide a principled understanding of how corpus size and LLM size interact, leaving the corpus-generator trade-off essentially unexplored. In particular, prior studies have not jointly examined retrieval corpus expansion and LLM size, leaving open the fundamental question of how corpus-generator trade-offs shape overall system performance.


%% file: 3-methodology.tex
\section{Methodology}
\label{sec:methodology}

We present a systematic framework to analyze corpus--generator trade-offs in RAG, asking when scaling the retrieval corpus can compensate for smaller LLMs and how this interaction varies across model sizes. Our design centers on two questions: (i) \textit{corpus scaling as compensation}---whether enlarging the corpus enables smaller generators to match or surpass larger ones; and (ii) \textit{differential effects across LLMs}---how benefits from corpus expansion change with model capacity.

\subsection{Retriever: Corpus Scaling}
Let $\mathcal{C}$ denote a fixed corpus. We simulate corpus scaling by randomly partitioning $\mathcal{C}$ into $N$ disjoint shards
$\{S_1,\ldots,S_N\}$ of approximately equal size:
\[
\Pi(\mathcal{C}) \;\to\; \{S_1, S_2, \ldots, S_N\},\quad
S_i \cap S_j = \varnothing \;\; \forall\, i \neq j,\quad 
\bigcup_{i=1}^{N} S_i = \mathcal{C}.
\]
A corpus scale $n \in \{1,\ldots,N\}$ is realized by activating $n$ shards; we use the canonical prefix $\mathcal{C}^{(n)} = \bigcup_{i=1}^{n} S_i$, so that $|\mathcal{C}^{(n)}| \propto n$. For a query $q$, the retriever operates on $\mathcal{C}^{(n)}$ to retrieve top-$k$ documents, segment them into chunks, rerank, and pass the top-$m$ chunks to the generator.

\subsection{Generator}
We consider a family of generators $\{M_x\}$, where $M_x$ denotes a model with parameter size $x$ drawn from the same architecture family. Each generator takes as input a fixed template consisting of the query and retrieved chunks.

\subsection{Trade-off Formalization}
We adopt a full-factorial design pairing each corpus scale $n \in \{1,\ldots,N\}$ with each generator $M_x$. Throughout, we keep retrieval, prompting, and decoding settings fixed, so that only corpus scale $n$ and model size $x$ vary. Let $P_m(n,x)$ denote the evaluation score under metric $m \in \{\text{F1}, \text{EM}\}$.

To quantify \emph{corpus-as-compensation}, we define
\[
n^{\star}(x_{small} \!\to\! x_{large}) \;:=\;
\min_{m \in \{\text{F1},\,\text{EM}\}}
\;\min \big\{ n \,:\, P_m(n, x_{small}) \,\ge\, P_m(1, x_{large}) \big\},
\]
the smallest corpus scale where a smaller generator $M_{x_{small}}$ matches the 1-shard baseline ($n=1$) of a larger generator $M_{x_{large}}$. We report $n^{\star}$ and efficiency curves across $(n,x)$ to characterize corpus--generator trade-offs. Section~\ref{sec:experiment} details datasets, metrics, and constants.

%% file: 4-Experiment.tex
\section{Experiment}
\label{sec:experiment}
Building on the methodology outlined in Section~\ref{sec:methodology}, we conducted a series of controlled experiments across different corpus scales to systematically evaluate our research questions.

\subsection{Benchmarks}
We evaluate on three open-domain QA benchmarks: 
NQ~\cite{kwiatkowski2019natural} (1,769 real Google queries from \textit{open-domain test set}), 
TriviaQA~\cite{joshi2017triviaqa} (1,000 encyclopedic questions randomly sampled from the 9.51k \textit{rc.web test split}), 
and WebQ~\cite{berant-etal-2013-semantic} (2,032 Google Suggest queries with Freebase annotations from \textit{standard test set}).

\noindent\textbf{Scope and limitation.}
Our evaluation targets open-domain QA with a large web corpus (ClueWeb22-A), which may overlap with modern LLM pretraining data; thus our results should be interpreted within this open-domain setting. Although our later analyses condition on questions that are initially incorrect without retrieval to isolate retrieval-driven gains, this setting still differs from domain-gap or enterprise RAG where the corpus is largely unseen during pretraining; thus, extrapolations should be made with caution.

\subsection{Evaluation Metrics}
We report Exact Match (EM) and token-level F1 using the official evaluation scripts. EM is the percentage of predictions that exactly match any normalized gold answer string. F1 is the average token-overlap F1 between the prediction and the gold answers, taking the maximum over gold answers for each question. Our analysis primarily focuses on these two metrics throughout the paper.

\subsection{Retriever: Implementation Details}
\textbf{Corpus and sharding.}
We use a 30\% subset of ClueWeb22-A~\cite{clueweb22}, comprising $\sim$264M English documents. The corpus is partitioned into 12 balanced shards of $\sim$22M documents each, via randomized local assignments to reduce topical skew (though some popularity bias may persist).

\noindent\textbf{Retrieval pipeline.}
We use \texttt{MiniCPM-Embedding-Light}~\cite{minicpm} for dense passage encoding and build ANN indices with \texttt{DiskANN}~\cite{diskann}, which supports fast multi-shard retrieval and has been used in other large-scale ClueWeb22-A retrievers~\cite{deepresearchgym}. As an auxiliary check, we also encoded the corpus with a similarly sized \texttt{Qwen3-Embedding-0.6B} encoder~\cite{qwen3embedding} and observed consistent corpus-scaling trends. For each corpus scale $n$, the retriever searches the active shards to retrieve the top-10 documents, segments them into overlapping chunks, and reranks the candidates; we then pass the top-8 chunks to the generator. For reranking, we use the higher-capacity \texttt{MiniCPM-Embedding} model from the same embedding family to improve ranking quality, while keeping all retrieval and reranking settings fixed across $n$ to isolate corpus scaling effects.



\subsection{Generator: Implementation Details}
We instantiate $\{M_x\}$ using the open-source Qwen3 base series~\cite{qwen3}: \texttt{Qwen3-0.6B}, \texttt{1.7B}, \texttt{4B}, \texttt{8B}, and \texttt{14B}. These models share the same tokenizer and a consistent architecture family as described in the Qwen3 technical report, spanning over an order of magnitude in parameter scale.

All models use identical prompting templates and decoding settings, enabling controlled comparisons of corpus--model trade-offs. While Qwen3 sizes share the same tokenizer and architecture family, different scales may involve minor training-recipe differences; thus, our results should be interpreted as empirical trends across sizes within the same Qwen3 series. We observed consistent trends in preliminary experiments with Qwen2.5 and early LLaMA models, but omit them due to the lack of a similarly homogeneous series and slower inference.

%% file: 5.1-corpus-compensation.tex
\subsection{The Effect of Corpus Scaling as Compensation}
We examine whether enlarging the retriever corpus can compensate for smaller LLMs, enabling them to match or surpass larger model. We parameterize corpus scale by the number of active shards $n$, where each shard indexes $\sim$22M documents from ClueWeb22-A; scaling the corpus corresponds to increasing $n$.

\begin{table}[H]
    \centering
    \caption{Natural Questions. Shaded cells mark the first scale where a smaller model catches up to the next model’s $n{=}1$ baseline, i.e., $n^{\star}(x_{small}\!\to\!x_{large})$.}
    \label{tab:nq-all-metrics-catchup}
    \renewcommand{\arraystretch}{0.95}
    \small
    \begin{threeparttable}
    \begin{tabularx}{\textwidth}{c *{10}{>{\centering\arraybackslash}X}}
    \toprule
    \textbf{Corpus} &
    \multicolumn{2}{c}{\textbf{$M_{0.6B}$}} &
    \multicolumn{2}{c}{\textbf{$M_{1.7B}$}} &
    \multicolumn{2}{c}{\textbf{$M_{4B}$}} &
    \multicolumn{2}{c}{\textbf{$M_{8B}$}} &
    \multicolumn{2}{c}{\textbf{$M_{14B}$}} \\
    \cmidrule(lr){2-3}\cmidrule(lr){4-5}\cmidrule(lr){6-7}\cmidrule(lr){8-9}\cmidrule(lr){10-11}
    \textbf{$\times n$} & F1 & EM & F1 & EM & F1 & EM & F1 & EM & F1 & EM \\
    \midrule
    1       & 25.37 & 16.68 & 33.44 & 23.74 & 40.27 & 28.49 & 42.42 & 30.24 & 44.26 & 32.50 \\
    $\times$2  & 29.90 & 19.50 & 38.85 & 28.32 & \cellcolor{catchup}\textbf{\textcolor{forestgreen}{44.16}} & \cellcolor{catchup}\textbf{\textcolor{forestgreen}{32.33}} & \cellcolor{catchup}\textbf{\textcolor{forestgreen}{45.59}} & \cellcolor{catchup}\textbf{\textcolor{forestgreen}{33.35}} & 46.96 & 34.65 \\
    $\times$3  & 30.99 & 20.75 & \cellcolor{catchup}\textbf{\textcolor{forestgreen}{41.00}} & \cellcolor{catchup}\textbf{\textcolor{forestgreen}{29.96}} & 46.00 & 34.31 & 46.88 & 34.71 & 48.94 & 36.69 \\
    $\times$4  & 32.72 & 21.54 & 41.08 & 29.85 & 46.33 & 34.43 & 48.16 & 35.56 & 48.98 & 37.08 \\
    $\times$5  & 33.26 & 22.84 & 41.49 & 30.19 & 46.49 & 34.37 & 48.40 & 35.39 & 49.51 & 37.14 \\
    $\times$6  & \cellcolor{catchup}\textbf{\textcolor{forestgreen}{34.00}} & 23.69 & 42.15 & 30.36 & 46.92 & 34.09 & 48.89 & 36.24 & 49.57 & 37.03 \\
    $\times$7  & 34.62 & \cellcolor{catchup}\textbf{\textcolor{forestgreen}{23.74}} & 42.15 & 30.92 & 47.25 & 34.77 & 48.04 & 35.44 & 49.54 & 36.80 \\
    $\times$8  & 34.72 & 24.19 & 42.12 & 31.09 & 46.91 & 34.60 & 48.04 & 35.73 & 49.34 & 36.80 \\
    $\times$9  & 34.49 & 23.97 & 42.04 & 30.75 & 46.64 & 34.31 & 48.27 & 35.61 & 49.36 & 36.63 \\
    $\times$10 & 35.35 & 24.59 & 42.86 & 31.32 & 47.08 & 34.43 & 48.62 & 36.07 & 49.75 & 37.37 \\
    $\times$11 & 35.48 & 25.10 & 42.78 & 30.98 & 47.43 & 34.71 & 49.06 & 36.01 & 50.38 & 37.59 \\
    $\times$12 & 35.56 & 25.10 & 43.14 & 31.32 & 47.89 & 34.88 & 48.73 & 35.78 & 50.43 & 37.70 \\
    \bottomrule
    \end{tabularx}
    \end{threeparttable}
\end{table}

We evaluate five Qwen3 variants ($M_{0.6B}$, $M_{1.7B}$, $M_{4B}$, $M_{8B}$, $M_{14B}$) and, for each model, vary $n$ by cumulatively activating shards under the same retrieval and prompting protocol, enabling controlled cross-model comparisons.

\subsubsection{Compensation Effect.}
\begin{figure}[H]
    \centering
    \begin{minipage}{0.49\textwidth}
        \includegraphics[width=\linewidth]{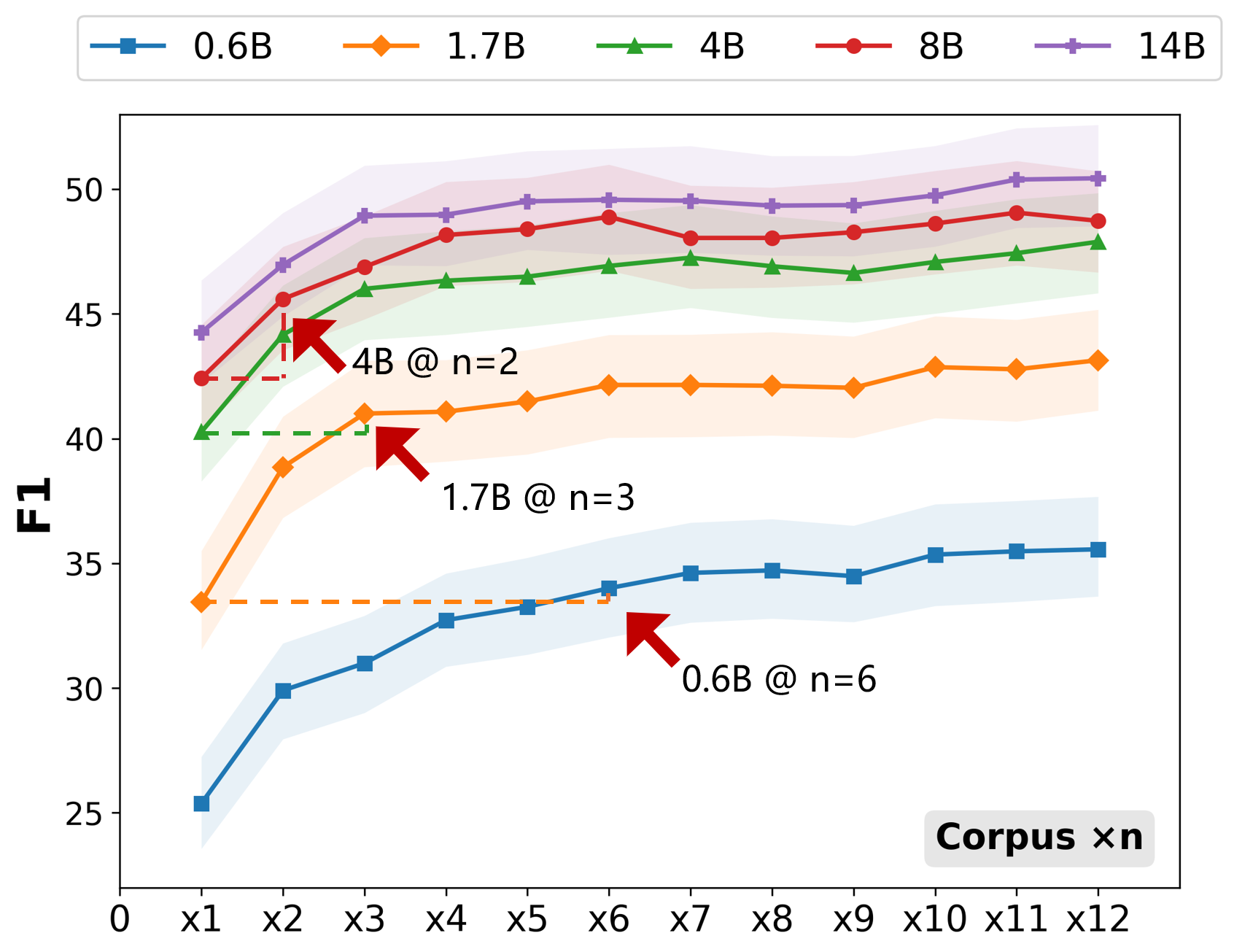}
        \caption{\textbf{F1} Gains from Scaling on NQ}
        \label{fig:nq-plot-f1}
    \end{minipage}
    \hfill
    \begin{minipage}{0.49\textwidth}
        \includegraphics[width=\linewidth]{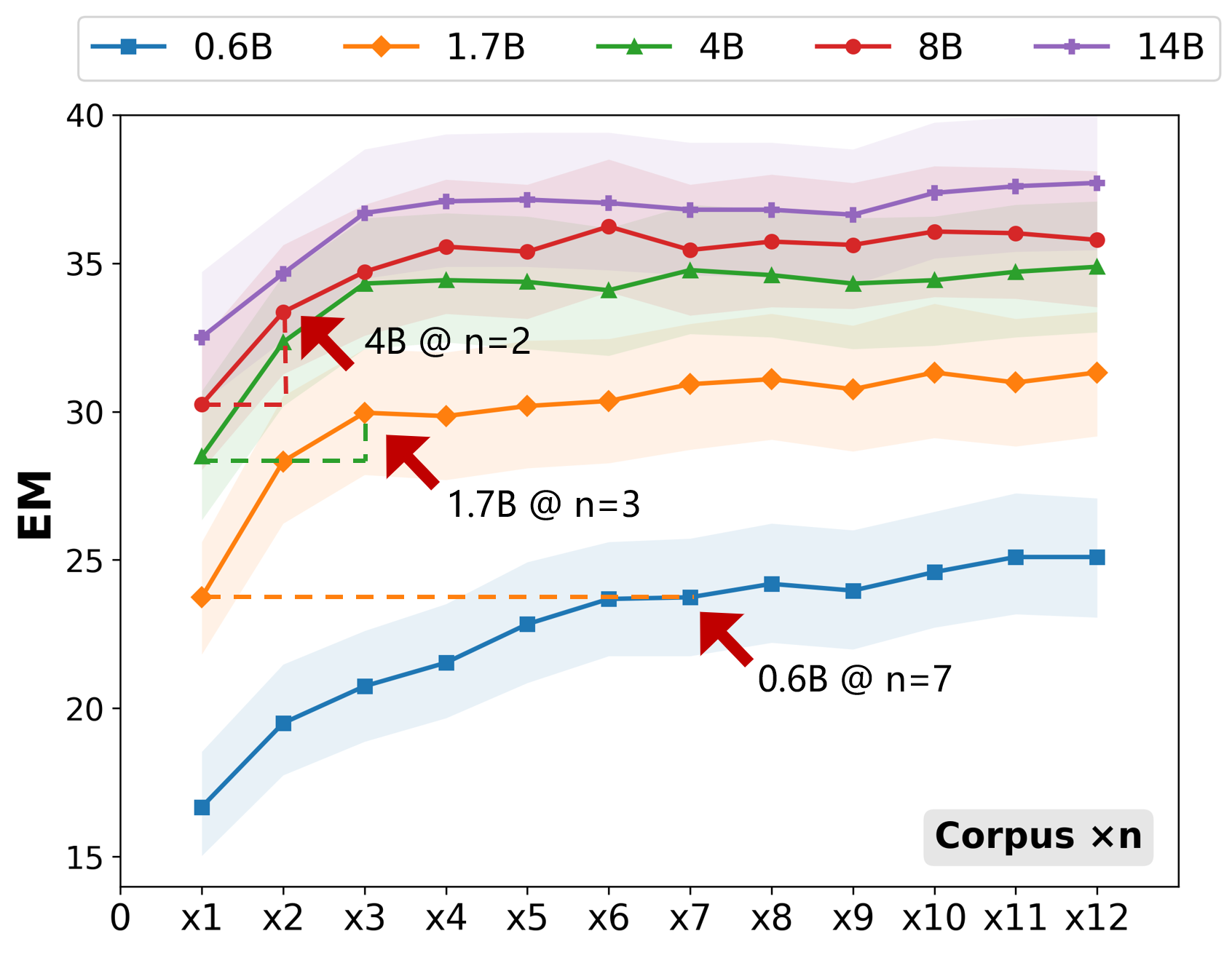}
        \caption{\textbf{EM} Gains from Scaling on NQ}
        \label{fig:nq-plot-em}
    \end{minipage}
\end{figure}

Our results show clear evidence that corpus expansion enables smaller models to match or even outperform larger counterparts. On NQ, as shown in Table~\ref{tab:nq-all-metrics-catchup}, we find that the smallest model needs more corpus to surpass the larger model
$n^{\star}(0.6\text{B}\!\to\!1.7\text{B})=6$. For larger models, it is much easier: $n^{\star}(4\text{B}\!\to\!8\text{B})=2$ and $n^{\star}(8\text{B}\!\to\!14\text{B})=2$. These results indicate that, under our setup, scaling corpus size can be a more effective and efficient lever than simply scaling LLM size. Figures~\ref{fig:nq-plot-f1} and~\ref{fig:nq-plot-em} visualize these catch-up points.

\begin{wraptable}{r}{0.56\textwidth}
    \centering
    \caption{$n^{\star}$ across datasets}
    \label{tab:catchup-summary}
    \renewcommand{\arraystretch}{1.2}
    \small
    \begin{tabular}{cccc|c}
        \toprule
        \textbf{$n^{\star}$} & \textbf{NQ} & \textbf{TriviaQA} & \textbf{WebQ} & \makecell{\textbf{NQ}\\{\tiny Qwen3-Emb}} \\
        \midrule
        0.6B$\rightarrow$1.7B & 6  & 10 & 9 & 5 \\
        1.7B$\rightarrow$4B   & 3  & 7  & 4 & 3 \\
        4B$\rightarrow$8B     & 2  & 2  & 3 & 2 \\
        8B$\rightarrow$14B    & 2  & 2  & 1 & 2 \\
        \bottomrule
    \end{tabular}
\end{wraptable}

The same trend holds on TriviaQA and WebQ (Table~\ref{tab:catchup-summary}). In the tiny-model regime (sub-2B parameters), corpus scaling is inefficient: e.g., $n^{\star}(0.6\text{B}\!\to\!1.7\text{B})=10$ and $n^{\star}(1.7\text{B}\!\to\!4\text{B})=7$ on TriviaQA, indicating weaker scaling efficiency at small model sizes. In contrast, once the generator reaches medium to large scale (roughly 4$\sim$14B parameters), \textbf{doubling} the corpus is typically sufficient to catch up with the next-tier model. For example, $n^{\star}(4\text{B} \to 8\text{B}) = 2$ and $n^{\star}(8\text{B} \to 14\text{B}) = 2$ on NQ and TriviaQA, and at most $n^{\star}=3$ on WebQ. Detailed results for TriviaQA and WebQ are provided in Table~\ref{tab:triviaqa-all-metrics-catchup} and Table~\ref{tab:webq-all-metrics-catchup} in the Appendix; using \texttt{Qwen3-Embedding} as the retriever on NQ yields nearly identical $n^{\star}$ patterns, suggesting this trend transfers across embedding models.

\subsubsection{Serving-Time Compute Proxy.}
We provide a simple \emph{serving-time} compute proxy by estimating FLOPs per query for retrieval and generation. Under a representative setting (top-$K{=}10$, $L{=}5K{=}50$, dot-product retrieval with $D{=}1024$; generation with $S{\approx}2048$ prompt tokens and $T{\approx}64$ output tokens), a single-shard DiskANN retrieval costs $\sim$7.168\,MFLOPs/query, while Qwen3-0.6B and Qwen3-1.7B generation cost $\sim$1.736\,TFLOPs/query and $\sim$5.953\,TFLOPs/query, respectively. This orders-of-magnitude gap is already sufficient to show that, in our setup, arithmetic compute is dominated by generation rather than retrieval. We focus on \emph{inference-time} compute because our experiments vary serving-side design choices such as corpus scaling and generator size. This proxy does not account for one-time costs such as LLM pretraining or index/shard construction.

\subsubsection{Corpus Quality vs.\ Quantity.}
\begin{figure}[htbp]
    \centering
    \begin{minipage}{0.5\textwidth}
        \includegraphics[width=\linewidth]{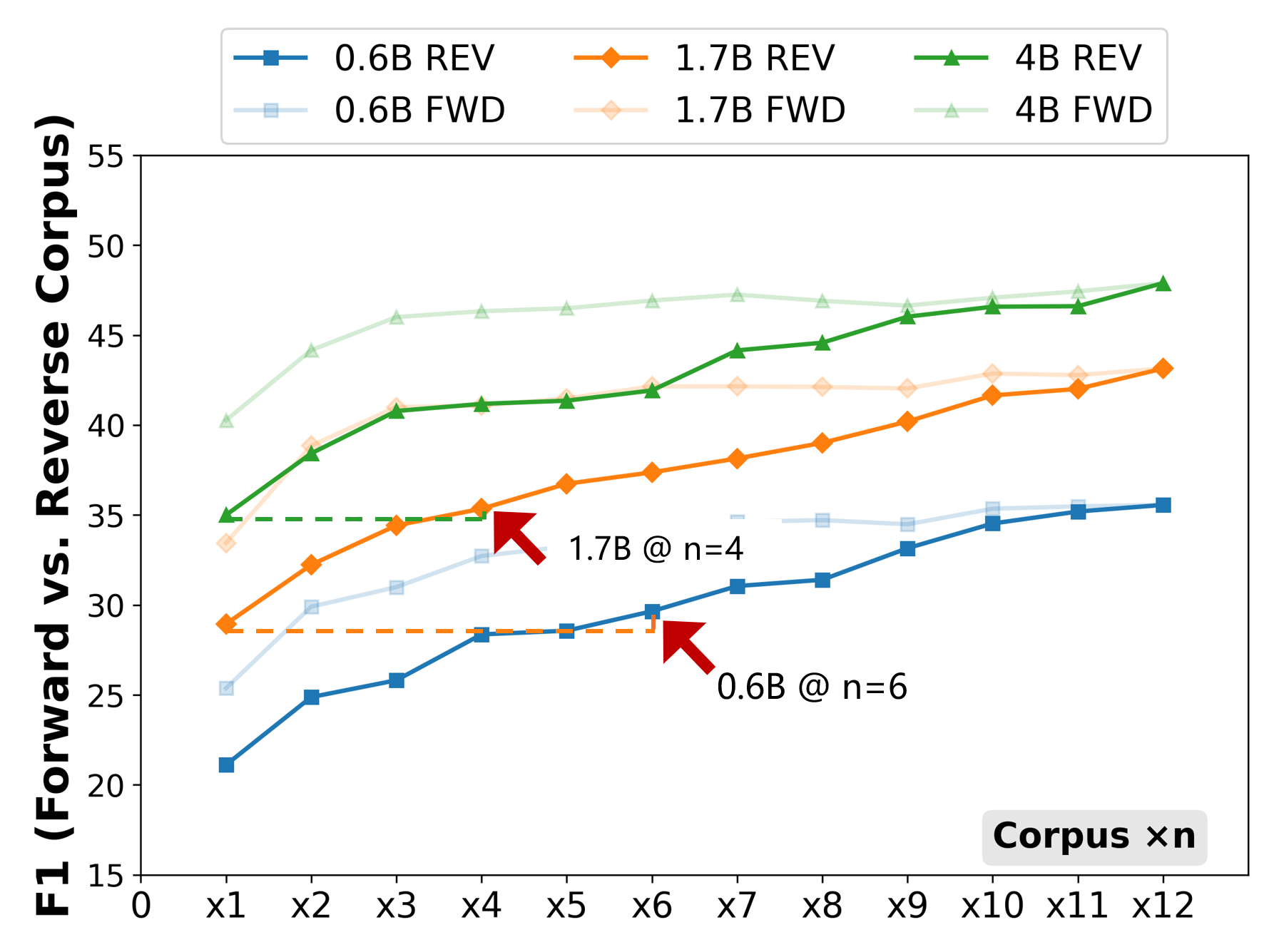}
        
    \end{minipage}
    \hfill
    \begin{minipage}{0.49\textwidth}
        \centering
        \setlength{\tabcolsep}{4pt}
        \renewcommand{\arraystretch}{1.8}
        \small
        \begin{tabular}{lcc}
            \toprule
            \textbf{$n^{\star}$} & NQ-FWD & \textbf{NQ-REV}  \\
            \midrule
            0.6B$\rightarrow$1.7B & 6  & 6 \\
            1.7B$\rightarrow$4B   & 3  & 4  \\
            4B$\rightarrow$8B     & 2  & 2  \\
            8B$\rightarrow$14B    & 2  & 2  \\
            \bottomrule
        \end{tabular}
    \end{minipage}
    \caption{F1 and Catch-up Thresholds under \textbf{Reversed} Corpus Scaling. Left: F1 when using forward (FWD) vs.\ reversed (REV) corpus scaling order. Right: corresponding catch-up thresholds.}
    \label{fig:nq-plot-f1-rev}
\end{figure}

Shards are balanced in size but not perfectly uniform, since randomization was applied locally rather than globally. Consequently, shard order may retain residual crawl/popularity gradients (e.g., earlier shards being slightly more likely to contain higher-traffic or higher-quality pages). To probe sensitivity to corpus quality, we reverse the shard order at retrieval time, replacing ${S_1,\ldots,S_n}$ with ${S_{N-n+1},\ldots,S_N}$, which delays the ``head'' content. As expected, this yields a modest drop in absolute end-task QA performance (Figure~\ref{fig:nq-plot-f1-rev}, left).

In other words, lower average corpus quality shifts performance downward, yet the \emph{relative} additional corpus needed for a smaller model to catch up with the next larger model remains largely stable. This supports our conclusion: corpus quality affects absolute accuracy, but scaling corpus quantity can still enable smaller generators to match or overtake larger ones with similar additional context. For consistency, we report all subsequent results using the \textbf{forward}  order.

\subsubsection{Why Does Corpus Scaling Improve RAG?}
At the micro level, corpus scaling increases the likelihood that retrieved passages explicitly contain the gold answer. With a small corpus scale ($n=1$), retrieved chunks often lack factual mentions. The larger $n$ brings the direct answer terms into the context, providing the generator with grounding evidence as intended in RAG.

\begin{tcolorbox}[title=Case Analysis,
    colback=white, colframe=black!80!white, boxrule=0.6pt,
    arc=2mm, left=6pt,right=6pt,top=4pt,bottom=4pt]
    \noindent
    \scriptsize
    \textbf{Question:} ``\textcolor{red}{Obey your thirst}'' is the advertising slogan for which soft drink?
    \textbf{Answer:} \hl{Sprite}
    
        \begin{tcolorbox}[title=Retrieved Fragment with Corpus scale \textnormal{$n = 4$},
        colback=black!5!white, colframe=black!80!white, boxrule=0.5pt,
        arc=2mm, left=4pt,right=4pt,top=6pt,bottom=6pt]
        \footnotesize
        \scriptsize
        \textbf{Retrieved passage:} \\
        ...\textcolor{red}{Obey Your Thirst} (Oct 1, 1997). Ever heard that catchy slogan for \hl{Sprite}?  
        ``Image is nothing. Thirst is everything. \textcolor{red}{Obey your thirst}.''  
        In the summer of 1996, Coca-Cola, who manufactures \hl{Sprite} products, was looking to change the image of its sparkling soda...''
        
        \end{tcolorbox}
\end{tcolorbox}

\begin{figure}[H]
    \centering
    \begin{minipage}{1.0\textwidth}
    \includegraphics[width=\linewidth]{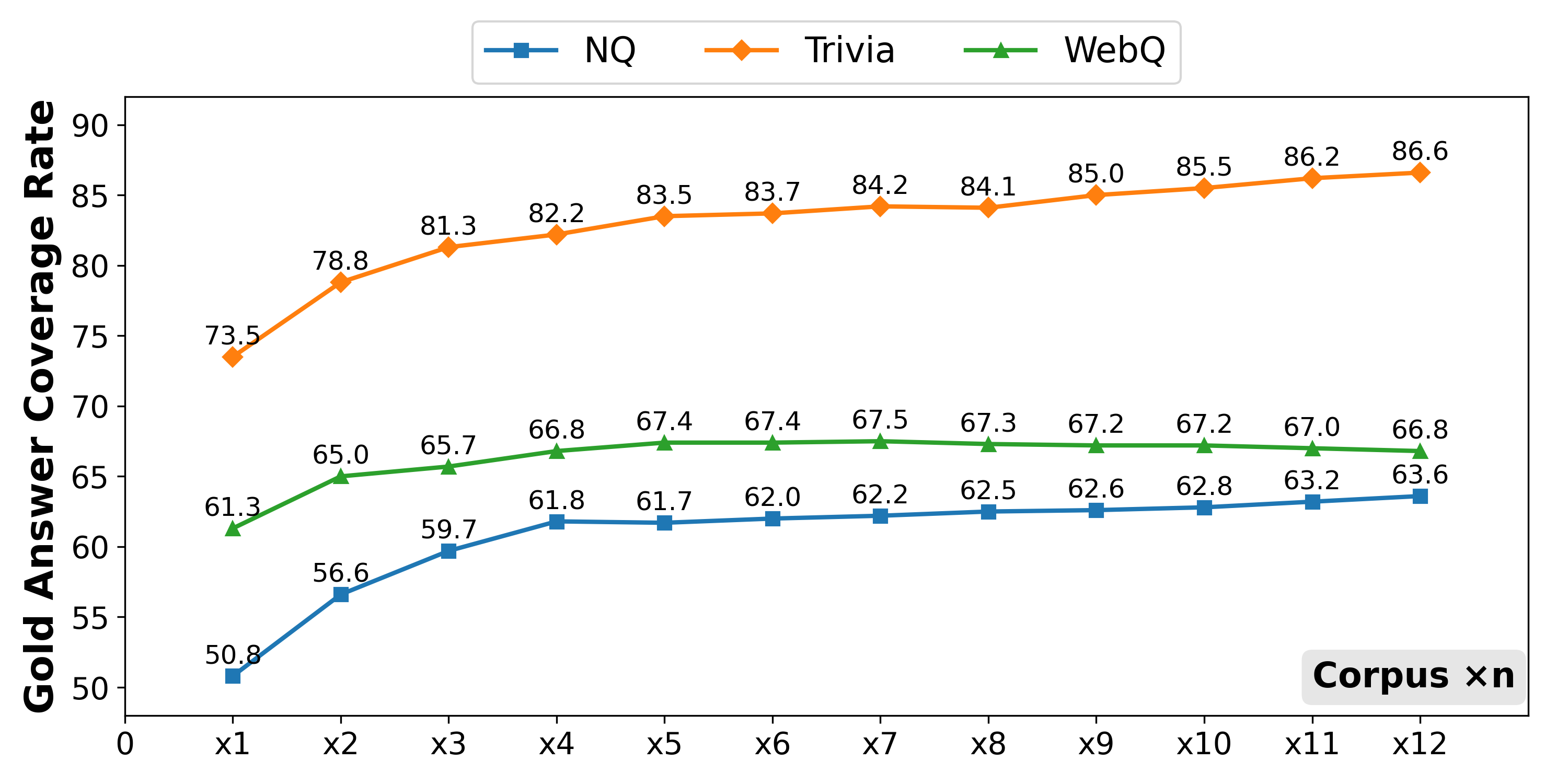}
    \caption{Gold Answer Coverage Rate for Forward Scaling}
    \label{fig:gold-coverage}
    \end{minipage}
\end{figure}

At the aggregate level, we measure the probability that at least one of the top-8 retrieved chunks fed into the generator contains a gold answer string, using the same normalization/aliasing as our EM metric. We refer to this probability as the \textbf{Gold Answer Coverage Rate}, which upper-bounds achievable EM under perfect reasoning. Figure~\ref{fig:gold-coverage} shows two key findings:
\begin{itemize}
    \item \textbf{Monotonic Growth.} Gold answer coverage often rises consistently with corpus scale, confirming that corpus expansion increases the likelihood of providing usable evidence.
    \item \textbf{Dataset Variation.} The magnitude of this benefit differs across benchmarks. TriviaQA exhibits substantially higher coverage than NQ or WebQ, indicating stronger overlap between its information needs and ClueWeb22.
\end{itemize}


%% file: 5.2-llm-sizes.tex
\subsection{Differential Effects Across LLM Sizes}

To analyze how retrieval corpus size differentially affects LLMs of varying scales, we focus on questions that are \emph{initially unanswerable} without retrieval and examine how performance changes as corpus size increases. Since correctness is most relevant here, we primarily rely on the Exact Match (EM) metric.

\subsubsection{Classification Methodology}

Let $n\in\{0,1,\dots,12\}$ denote the corpus size in shards, where $n{=}0$ represents the no-retrieval baseline.
We define the \textbf{Context-Benefited Success Rate (CB) at shard $n$} as
\[
\mathrm{CB}@n
\;:=\;
\Pr\!\big(EM_{n\text{-shard}}=1 \,\big|\, EM_{0\text{-shard}}=0\big)
\]
i.e., the \emph{empirical proportion} of initially unanswerable questions that become answerable once $n$ shards are available.
By construction, $\mathrm{CB}@0=0$. For $n\ge 1$, we also report the marginal improvement.
\[
\Delta_n \;:=\; \mathrm{CB}@n \;-\; \mathrm{CB}@{(n-1)}
\]
which captures the fraction of initially unanswerable questions solved at shard $n$. \emph{Note} that although $\mathrm{CB}@n$ is computed cumulatively, it is an empirical statistic and need not be strictly monotonic in $n$ (e.g., additional shards may introduce noise).

Although $\mathrm{CB}@n$ reflects the gains realized, it is bounded above by \emph{Gold Answer Coverage Rate} at shard $n$.
We define the \textbf{Utilization Ratio} as
\[
\mathrm{Ratio}@n \;:=\; \frac{\mathrm{CB}@n}{\mathrm{Coverage}@n}
\]

This ratio quantifies an LLM’s ability to take advantage of the retrieved evidence:
$\mathrm{Coverage}@n$ indicates how often the gold answer is retrievable, 
while $\mathrm{CB}@n$ records how often the model succeeds when given the opportunity.

\begin{figure}[H]
    \centering
    \begin{minipage}{0.49\textwidth}
        \includegraphics[width=\linewidth]{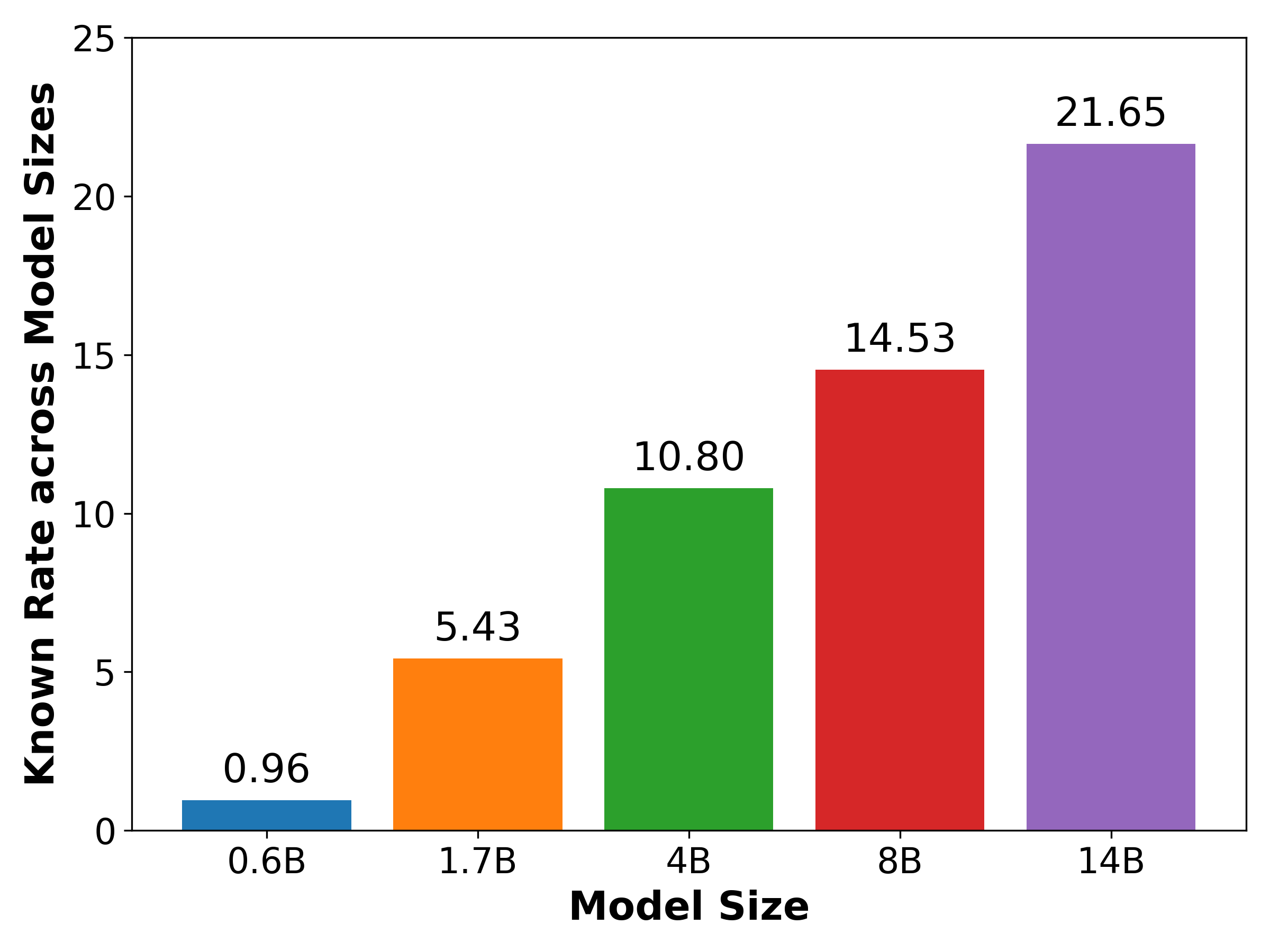}
        \caption{Known Rate on NQ}
        \label{fig:known-rate}
    \end{minipage}
    \hfill
    \begin{minipage}{0.49\textwidth}
        \includegraphics[width=\linewidth]{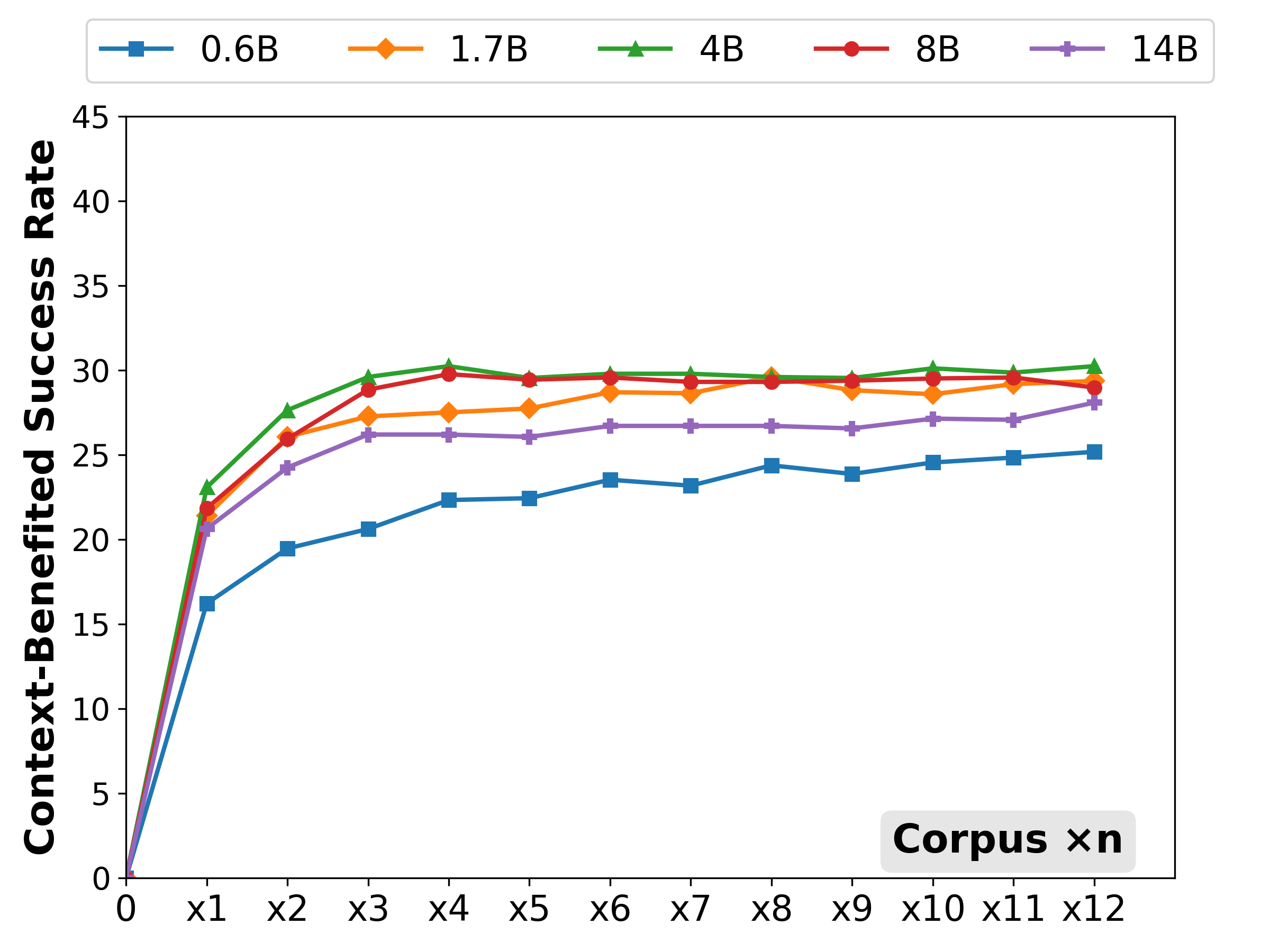}
        \caption{CB Rate on NQ (FWD)}
        \label{fig:nq-cb-rate}
    \end{minipage}
\end{figure}

CB excludes questions already correct at \(n=0\) (Known), so it isolates retrieval-driven gains by removing parametric knowledge effects. The \textbf{Known Rate} in Figure~\ref{fig:known-rate} summarizes how much each model can answer without retrieval.

Figure~\ref{fig:nq-cb-rate} reveals a consistent and model-invariant pattern for how the scaling helps answer questions that are \emph{initially unanswerable} without context, with similar trends observed on TriviaQA and WebQ (see the Appendix).

\subsubsection{Initial Jump, Subsequent Growth, and Saturation.}

\paragraph{The Critical Impact of Initial Retrieval.}
The most dramatic performance jump occurs when moving from zero context to a single shard across models, with $\Delta_1$ ranging from 16.2\% to 20.6\%, whereas $\Delta_2$ ranges from only 2.8\% to 4.4\% (Figure~\ref{fig:nq-cb-rate}). This dominance of initial retrieval persists even with low-quality corpora: for $M_{1.7B}$ with reversed shard ordering, \(\Delta_1=16.6\%\) versus \(\Delta_2=2.6\%\). This highlights the primary benefit of RAG: even a small corpus immediately fills a substantial fraction of knowledge gaps.

\begin{figure}[H]
    \centering
    \begin{minipage}{1.0\textwidth}
    \includegraphics[width=\linewidth]{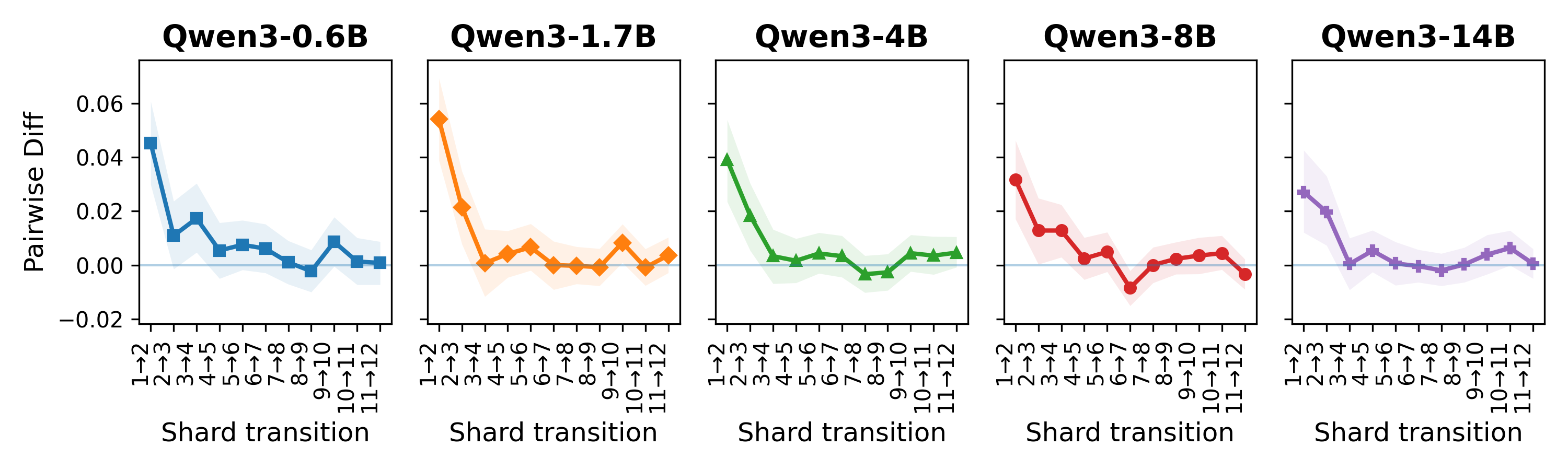}
    \caption{\textbf{F1} Pairwise Bootstrap on NQ}
    \label{fig:nq-plot-pairwise-boost}
    \end{minipage}
\end{figure}

\paragraph{Model-invariant growth and saturation.}
Across all LLM sizes, adjacent-shard improvements in downstream QA (F1) exhibit a highly consistent pattern. The paired bootstrap differences from shard $n$ to $n{+}1$ show large, reliably positive gains in the early steps, followed by rapid decay toward a near-zero regime (Figure~\ref{fig:nq-plot-pairwise-boost}). For all model sizes, the 95\% CI typically begins to include zero around the $3{\to}4$ or $4{\to}5$ transition, indicating diminishing returns and that later fluctuations are often within statistical noise rather than systematic improvement; we observe the same conclusion when computing paired differences using EM.

\begin{figure}[H]
    \centering
    \begin{minipage}{1.0\textwidth}
    \includegraphics[width=\linewidth]{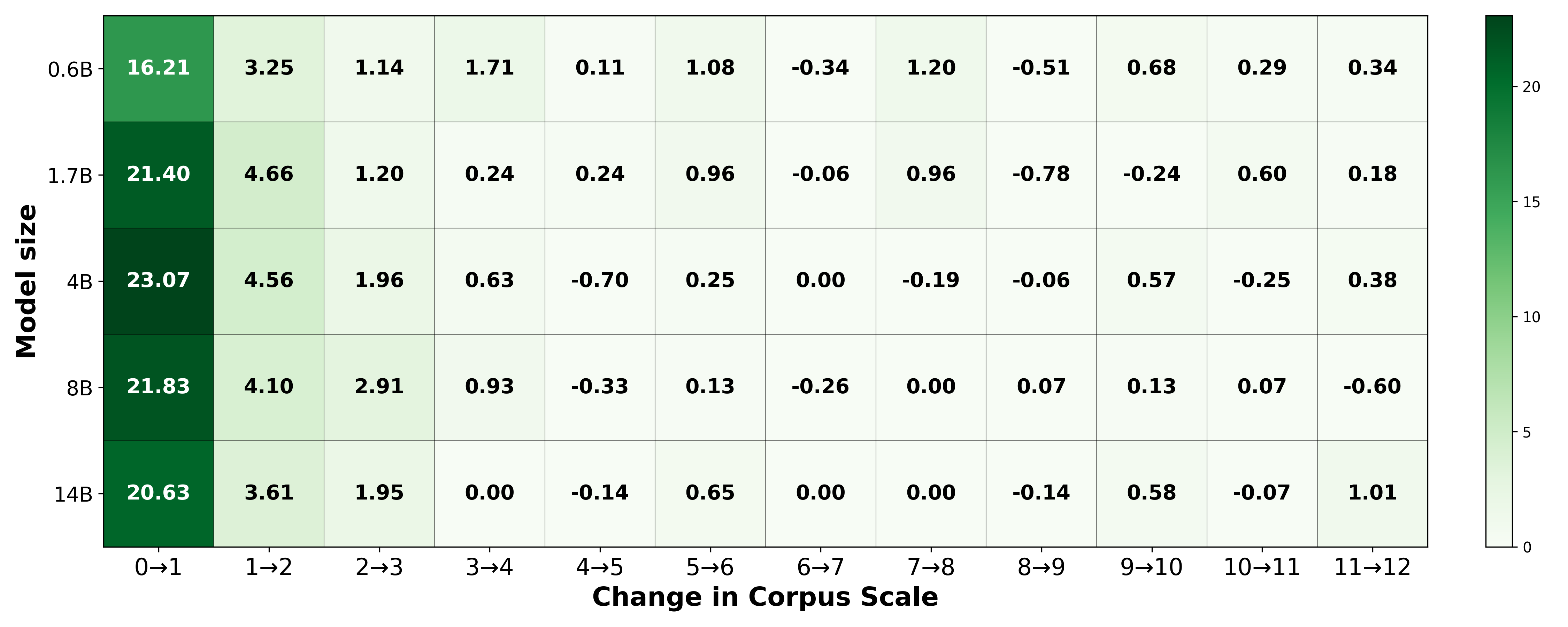}
    \caption{Per-shard CB gains $\Delta_n$ on NQ (FWD)}
    \label{fig:cb-deltas}
    \end{minipage}
\end{figure}

A consistent picture emerges for retrieval-driven gains on initially unanswerable questions. Corpus scaling yields a qualitatively similar CB trajectory across models: a sharp first jump, sustained gains up to roughly $n\!\approx\!6$, and diminishing returns thereafter. The per-shard increments $\Delta_n$ also follow a nearly identical pattern across models: peaking early and tapering to near zero (Figure~\ref{fig:cb-deltas}). Together, these results suggest a size-invariant retrieval effect: additional shards do not yield systematically larger marginal gains for larger generators than for smaller ones. In practice, corpus expansion mainly shifts performance upward without materially changing the shape of the growth curve.

\subsubsection{LLM Context Utilization Remains Stable Across Corpus Scales.}
\begin{figure}[H]
    \centering
    \begin{minipage}{0.8\textwidth}
    \includegraphics[width=\linewidth]{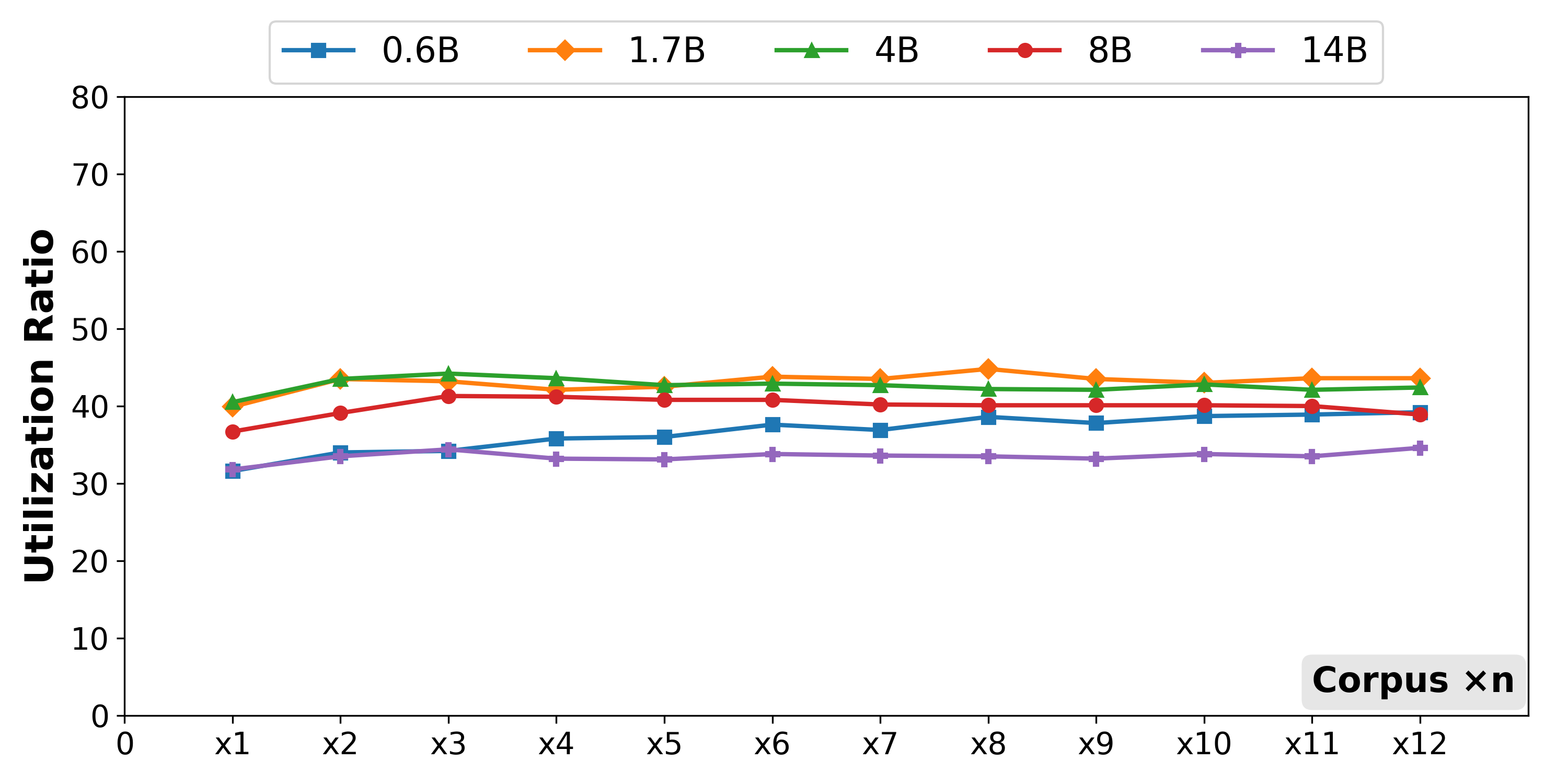}
    \caption{Utilization Ratio across models on NQ (FWD)}
    \label{fig:rate-progressive}
    \end{minipage}
\end{figure}

As shown in Figure~\ref{fig:rate-progressive}, the Utilization Ratio stays approximately constant across corpus scales and varies only slightly across models. Although both $\mathrm{CB}@n$ and $\mathrm{Coverage}@n$ increase with $n$, their ratio remains stable, indicating that corpus scaling primarily improves answer-bearing evidence coverage rather than changing how efficiently generators use the provided context. As a result, gains largely track the increased availability of relevant evidence as the corpus grows.

\paragraph{Non-monotonic Context Utilization}
One might expect larger LLMs to always leverage retrieved context more effectively, but Figure~\ref{fig:rate-progressive} shows otherwise. Mid-sized models ($M_{1.7B}$ and $M_{4B}$) achieve the highest Utilization Ratio (peaking near 42\%), while the largest $M_{14B}$ lags behind. This suggests that context utilization does not grow monotonically with model size, and mid-sized models can sometimes exploit retrieval more efficiently than their larger counterparts.


%% file: 6-conclusion.tex
\section{Conclusion}

In this paper, we asked whether scaling the retrieval corpus can substitute for scaling the generator in RAG, and how corpus size interacts with model size under a fixed evidence budget. Using controlled evaluations on NQ, TriviaQA, and WebQ with standardized prompting and context formatting, we characterize a corpus--generator trade-off while holding the presented evidence constant.\\

Our results indicate that, in this open-domain setting, corpus scaling can often \emph{partially offset} model downsizing. Across datasets, enlarging the corpus is a reliable lever: smaller or mid-sized generators paired with larger corpora frequently approach, and sometimes exceed, the accuracy of larger models under the same evidence budget. In several settings, moving up corpus tiers closes the gap of one to two model-size tiers, suggesting that increasing corpus scale can in many cases yield gains comparable to moving to a larger generator tier.\\

A consistent mechanism explains these gains: \emph{improvements are primarily driven by increased coverage of answer-bearing evidence rather than changes in utilization efficiency}. As the corpus grows, the likelihood that retrieved passages contain the gold answer increases consistently, whereas the model's context-utilization ratio remains roughly stable across shard counts and model sizes; we observe the same qualitative corpus-scaling trends under an auxiliary retriever built with a different encoder, suggesting the main findings are not specific to a single embedding model. Thus, scaling mainly works by raising the hit rate of relevant evidence, instead of altering how effectively models exploit the provided context. At the same time, both the paired bootstrap differences between adjacent shard counts and the per-shard CB increments $\Delta_n$ show that marginal gains rapidly decay and become indistinguishable from zero beyond roughly a $5\times$ corpus increase, indicating clear diminishing returns at larger scales.\\

From a practical perspective, our findings suggest that retrieval-side scaling is an important design axis when additional corpus is available and retrieval costs are acceptable under deployment constraints. Under our evidence budget, a simple serving-time FLOPs proxy indicates an orders-of-magnitude compute gap between single-shard retrieval and generation, suggesting that overall arithmetic compute is typically dominated by the generator (notwithstanding retrieval's memory/IO and indexing overheads). In our experiments, mid-sized generators tend to convert improved evidence coverage into end-task gains more effectively than very small models (which require steep corpus expansions) and very large models (which yield smaller marginal gains). While we mainly focus on the Qwen3 family due to the lack of other open-source LLM series with homogeneous, wide-ranging variants, we hope to extend this analysis to additional model and retriever families, as well as more domain-specific and siloed corpora, in future work. Finally, tracking \emph{gold-answer coverage} together with the \emph{Utilization Ratio} provides practical diagnostics for when increasing retrieval scale is likely to translate into downstream improvements; when corpus expansion is limited (e.g., enterprise or specialized domains), the same evidence-centric view suggests a \emph{coverage-first} strategy via domain-focused curation and retrieval quality improvements, rather than adding low-yield documents.

%% file: appendix.tex
\section*{Appendix}

\begin{table}[H]
    \centering
    \caption{$n^{\star}(x_{small}\!\to\!x_{large})$ for TriviaQA.}
    \label{tab:triviaqa-all-metrics-catchup}
    \renewcommand{\arraystretch}{0.9}
    \scriptsize
    \begin{threeparttable}
    \begin{tabularx}{\textwidth}{c *{10}{>{\centering\arraybackslash}X}}
    \toprule
    \textbf{Corpus} &
    \multicolumn{2}{c}{\textbf{$M_{0.6B}$}} &
    \multicolumn{2}{c}{\textbf{$M_{1.7B}$}} &
    \multicolumn{2}{c}{\textbf{$M_{4B}$}} &
    \multicolumn{2}{c}{\textbf{$M_{8B}$}} &
    \multicolumn{2}{c}{\textbf{$M_{14B}$}} \\
    \cmidrule(lr){2-3}\cmidrule(lr){4-5}\cmidrule(lr){6-7}\cmidrule(lr){8-9}\cmidrule(lr){10-11}
    \textbf{$\times n$} & F1 & EM & F1 & EM & F1 & EM & F1 & EM & F1 & EM \\
    \midrule
    1  & 44.74 & 37.20 & 57.40 & 49.80 & 66.46 & 59.80 & 69.86 & 62.90 & 73.16 & 66.60 \\
    $\times$2  & 49.73 & 41.00 & 62.68 & 55.10 & \cellcolor{catchup}\textbf{\textcolor{forestgreen}{73.17}} & \cellcolor{catchup}\textbf{\textcolor{forestgreen}{65.50}} & \cellcolor{catchup}\textbf{\textcolor{forestgreen}{76.32}} & \cellcolor{catchup}\textbf{\textcolor{forestgreen}{68.80}} & 77.72 & 71.10 \\
    $\times$3  & 51.86 & 43.60 & 64.90 & 57.20 & 75.99 & 68.50 & 77.59 & 69.90 & 79.59 & 72.50 \\
    $\times$4  & 53.20 & 45.10 & 66.18 & 59.00 & 76.56 & 69.60 & 78.04 & 71.20 & 80.75 & 74.00 \\
    $\times$5  & 54.79 & 46.40 & 66.08 & 59.30 & 76.86 & 69.40 & 78.00 & 71.00 & 80.68 & 73.90 \\
    $\times$6  & 55.26 & 47.55 & 66.01 & 58.16 & 76.60 & 69.77 & 77.58 & 70.57 & 80.70 & 74.07 \\
    $\times$7  & 55.59 & 47.30 & \cellcolor{catchup}\textbf{\textcolor{forestgreen}{66.89}} & 59.30 & 76.68 & 69.80 & 78.11 & 70.90 & 80.55 & 73.90 \\
    $\times$8  & 56.52 & 48.00 & 67.31 & 59.50 & 77.73 & 69.40 & 78.27 & 71.30 & 80.77 & 74.20 \\
    $\times$9  & 55.62 & 47.20 & 68.32 & \cellcolor{catchup}\textbf{\textcolor{forestgreen}{60.90}} & 77.61 & 70.40 & 79.30 & 72.50 & 81.10 & 74.40 \\
    $\times$10 & \cellcolor{catchup}\textbf{\textcolor{forestgreen}{57.61}} & 48.90 & 68.23 & 60.90 & 77.92 & 70.90 & 79.88 & 73.30 & 81.56 & 74.90 \\
    $\times$11 & 58.74 & \cellcolor{catchup}\textbf{\textcolor{forestgreen}{50.40}} & 68.60 & 61.40 & 78.13 & 71.00 & 79.68 & 72.90 & 82.13 & 75.50 \\
    $\times$12 & 57.86 & 49.05 & 68.44 & 61.16 & 77.89 & 70.77 & 80.05 & 73.27 & 82.00 & 75.28 \\
    \bottomrule
    \end{tabularx}
    \end{threeparttable}
\end{table}

\begin{table}[H]
    \centering
    \caption{$n^{\star}(x_{small}\!\to\!x_{large})$ for WebQuestions.}
    \label{tab:webq-all-metrics-catchup}
    \renewcommand{\arraystretch}{0.9}
    \scriptsize
    \begin{threeparttable}
    \begin{tabularx}{\textwidth}{c *{10}{>{\centering\arraybackslash}X}}
    \toprule
    \textbf{Corpus} &
    \multicolumn{2}{c}{\textbf{$M_{0.6B}$}} &
    \multicolumn{2}{c}{\textbf{$M_{1.7B}$}} &
    \multicolumn{2}{c}{\textbf{$M_{4B}$}} &
    \multicolumn{2}{c}{\textbf{$M_{8B}$}} &
    \multicolumn{2}{c}{\textbf{$M_{14B}$}} \\
    \cmidrule(lr){2-3}\cmidrule(lr){4-5}\cmidrule(lr){6-7}\cmidrule(lr){8-9}\cmidrule(lr){10-11}
    \textbf{$\times n$} & F1 & EM & F1 & EM & F1 & EM & F1 & EM & F1 & EM \\
    \midrule
    1  & 27.63 & 14.81 & 33.91 & 18.01 & 37.01 & 20.28 & 38.32 & \cellcolor{catchup}\textbf{\textcolor{forestgreen}{21.70}} & 38.68 & 21.65 \\
    $\times$2  & 28.95 & 15.70 & 35.40 & 19.09 & 38.20 & 20.92 & \cellcolor{catchup}\textbf{\textcolor{forestgreen}{38.92}} & 21.99 & 39.46 & 22.19 \\
    $\times$3  & 29.99 & 16.09 & 35.41 & 19.64 & \cellcolor{catchup}\textbf{\textcolor{forestgreen}{38.81}} & 20.96 & 39.44 & 22.60 & 40.49 & 22.93 \\
    $\times$4  & 31.37 & 17.62 & 35.96 & \cellcolor{catchup}\textbf{\textcolor{forestgreen}{20.28}} & 39.47 & 21.60 & 40.17 & 23.52 & 41.32 & 24.02 \\
    $\times$5  & 31.68 & 17.66 & 36.35 & 20.28 & 39.59 & 21.55 & 40.29 & 23.38 & 41.08 & 23.77 \\
    $\times$6  & 31.58 & 17.32 & 36.47 & 20.08 & 39.98 & 21.66 & 40.24 & 23.13 & 41.22 & 23.67 \\
    $\times$7  & 31.37 & 17.32 & 36.46 & 20.03 & 39.65 & 21.65 & 40.12 & 22.63 & 41.25 & 23.52 \\
    $\times$8  & 31.80 & 17.62 & 36.64 & 19.92 & 39.62 & 21.57 & 40.00 & 22.93 & 40.84 & 23.08 \\
    $\times$9  & 32.33 & \cellcolor{catchup}\textbf{\textcolor{forestgreen}{18.09}} & 36.62 & 20.34 & 39.36 & 21.45 & 39.79 & 22.83 & 40.57 & 22.74 \\
    $\times$10 & 32.18 & 18.09 & 36.83 & 20.77 & 38.99 & 21.62 & 39.95 & 23.50 & 40.79 & 22.93 \\
    $\times$11 & 31.98 & 17.91 & 36.61 & 20.57 & 39.20 & 21.51 & 39.94 & 22.58 & 40.58 & 22.88 \\
    $\times$12 & 32.00 & 17.82 & 36.94 & 21.01 & 39.41 & 21.46 & 40.12 & 22.93 & 40.69 & 23.18 \\
    \bottomrule
    \end{tabularx}
    \end{threeparttable}
\end{table}


\begin{figure}[H]
    \centering
    \begin{minipage}{0.49\textwidth}
        
        \includegraphics[width=\linewidth]{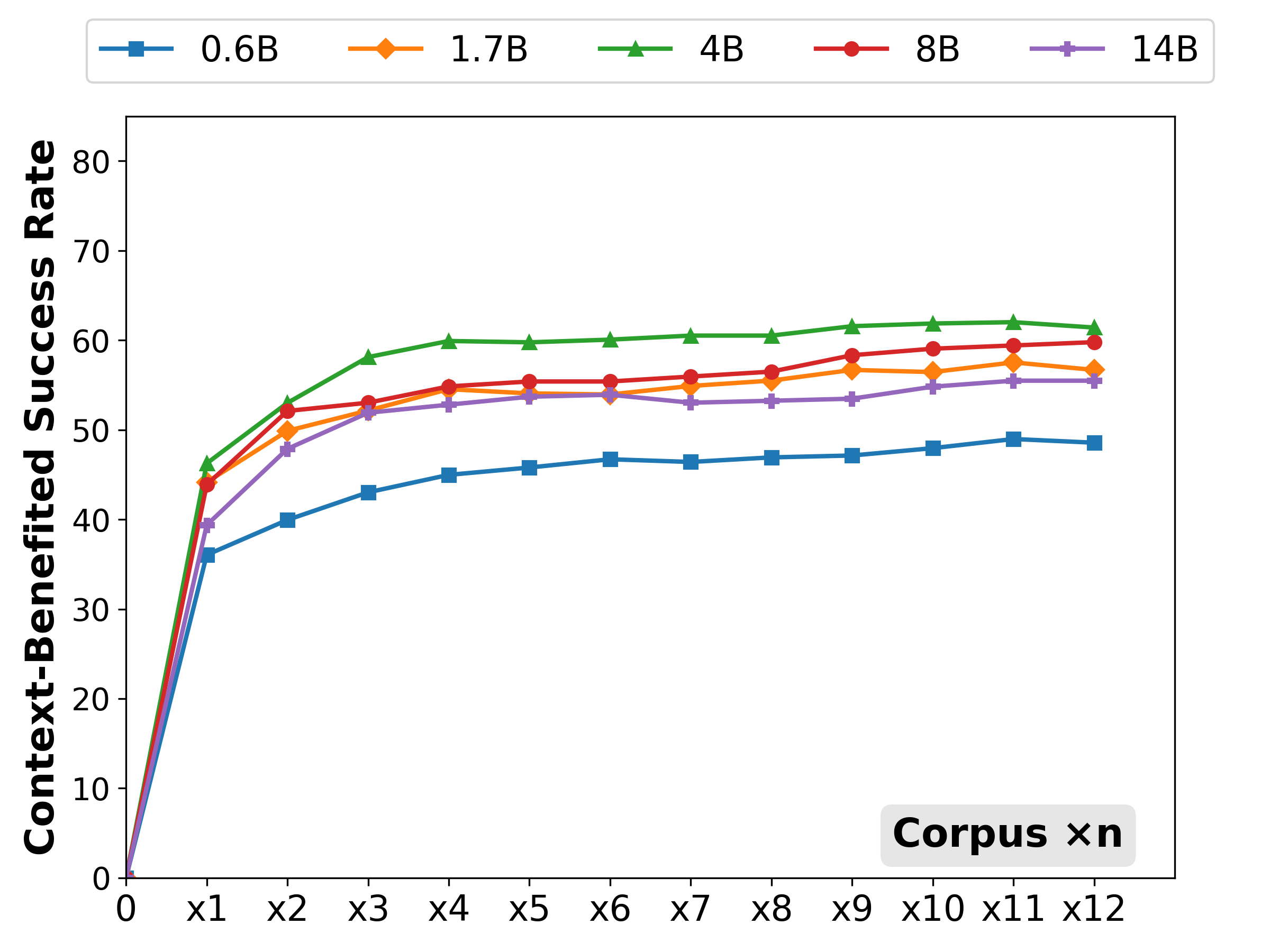}
        \caption{CB Rate for \textbf{TriviaQA}}
        \label{fig:trivia-cb-rate}
    \end{minipage}
    \hfill
    \begin{minipage}{0.49\textwidth}
       
        \includegraphics[width=\linewidth]{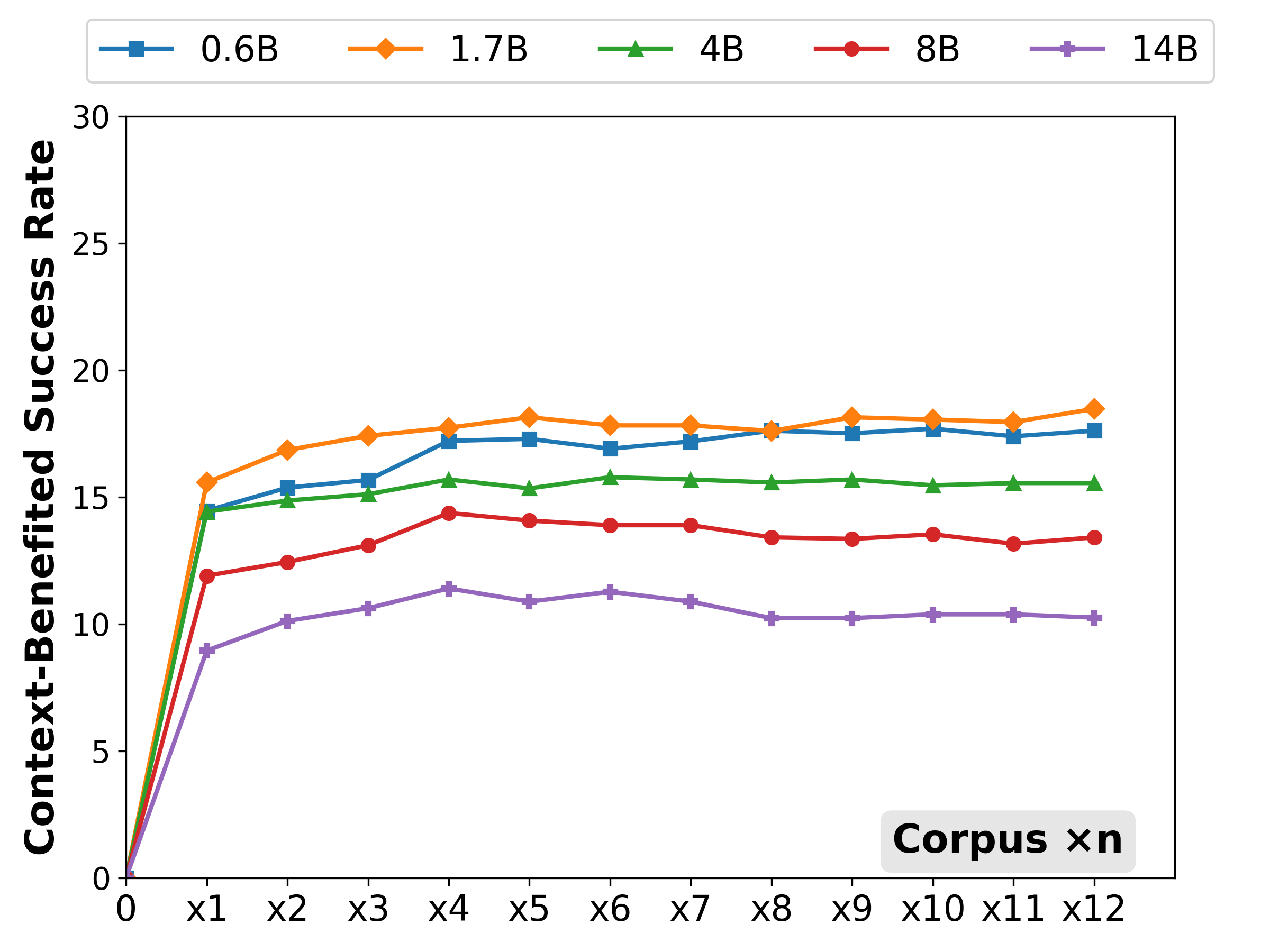}
        \caption{CB Rate for \textbf{WebQ}}
         \label{fig:webq-cb-rate}
    \end{minipage}
\end{figure}